# a net.be

# A Datamining Approach to the Short Title Catalogue Flanders: the Case of Early Modern Quiring Practices ¶

author: Tom Deneire (http://anet.be (http://anet.be)), last revision: 15 November 2018; keywords: datamining, STCV, handpress books, quiring, SQLite, Python, Pandas

This notebook contains a data mining approach to the Short Title Catalogue Flanders (http://www.stcv.be/ (http://www.stcv.be/)), which aims to record all books printed in Flanders up to 1801 (24.850 editions, per 31/08/2018). More specifically, it aims to analyse the Early Modern practice of 'quiring' gatherings in handpress book production. Gaskell's manual on bibliography defines quiring as follows:

*It was normal in the early days of printing to impose [sc. sheets] for gatherings of several sheets tucked, or quired, inside each other. Thus a folio gathering might consist of three folio sheets [i.e. three sheets folded once], the outermost of which contained pages 1 and 12 (printed from the outer forme) and pages 2 and 11 (from the inner forme); the middle sheet had pages 3 and 10, 4 and 9; and the innermost sheet had pages 5 and 8, 6 and 7. All three sheets were signed with the same letter (A1 on page 1, A2 on page 3, and A3 on page 5), and the folding is designated '2° in 6s'* (Philip Gaskell, A New Introduction to Bibliography. The Classic Manual of Bibliography (Oxford: OUP, 2012: first ed. 1972, second ed. 1995), p. 82.)

This research is done starting from SQLite export of the STCV data, embedded in Python, which allows for the data to be transferred to a Pandas dataframe. The results are then analysed with Pandas and plotted with Matlib.

## A. Extracting the data

The first step is to extract the necessary data for this quiring research with an SQLite query. This query contains a subquery to render the bibliographical format for each row of collation data. It also performs some data cleaning to remove unusable collation rows (e.g. `16 volumes`). We write this data directly to a Pandas dataframe and then offer some basic information about the dataframe and render the first ten rows. Next, we also save the data to a .csv file (`STCV_all.csv`) for browsing.

```python
import sqlite3
conn = sqlite3.connect('stcv.sqlite') #change this to your file location
cur = conn.cursor()
query = "select distinct collation.cloi, format, collation.collation_ka, impressum.impr
essum_ju1sv, impressum.impressum_ju2sv, impressum.impressum_pc, impressum.impressum_pl,
 impressum.impressum_uc, impressum.impressum_ug from collation join (select distinct co
llation.cloi as id, collation.collation_fm as format from collation) on id = collation.
cloi join impressum on impressum.cloi = collation.cloi where not format='' and not form
at='-' and not collation.collation_ka='' and not collation.collation_ka like '%#' and n
ot collation.collation_ka like '% volumes'"
data = []
cur.execute(query)
for x in cur.fetchall():
    try:
        data.append(x)
    except:
        print('Error:' + repr(x))
conn.close()

import pandas as pd
df1 = pd.DataFrame(data, columns=['identifier', 'format', 'collation', 'year1', 'year2'
, 'place_code', 'place_name', 'publisher_code', 'publisher_name'])
#The bottom row of this dataframe consists of the original column labels, which would i
nfere with future data analysis.
#Therefore we delete this row.
df1 = df1[:-1]
display(df1.info(), df1[0:11])
df1.to_csv('STCV_all.csv') #an encoding issue causes <pi> and <chi> characters to be re
ndered strangely
```

```
<class 'pandas.core.frame.DataFrame'>
RangeIndex: 28292 entries, 0 to 28291
Data columns (total 9 columns):
identifier       28292 non-null object
format           28292 non-null object
collation        28292 non-null object
year1            28292 non-null object
year2            28292 non-null object
place_code       28292 non-null object
place_name       28292 non-null object
publisher_code   28292 non-null object
publisher_name   28292 non-null object
dtypes: object(9)
memory usage: 994.7+ KB

None
```

| | identifier | format | collation | year1 | year2 | place_code | place_name | p |
|---|---|---|---|---|---|---|---|---|
| 0 | c:stcv:12840621 | octavo | 1# π<sup>4</sup> A-Z<sup>8</sup> 2A<sup>6</sup> | 1787 | 1789 | a::91.493.8000:1.13 | Brugge | |
| 1 | c:stcv:12840621 | octavo | 2# [A]<sup>8</sup> B-2A<sup>8</sup> 2B<sup>4</... | 1787 | 1789 | a::91.493.8000:1.13 | Brugge | |
| 2 | c:stcv:12840621 | octavo | 3# [A]<sup>8</sup> B-Z<sup>8</sup> (Z8 blank) | 1787 | 1789 | a::91.493.8000:1.13 | Brugge | |
| 3 | c:stcv:12840621 | octavo | 4# [A]<sup>8</sup> B-L<sup>8</sup> M<sup>4</su... | 1787 | 1789 | a::91.493.8000:1.13 | Brugge | |
| 4 | c:stcv:12840621 | octavo | 5# [A]<sup>8</sup> B-L<sup>8</sup> M<sup>6</su... | 1787 | 1789 | a::91.493.8000:1.13 | Brugge | |
| 5 | c:stcv:12840621 | octavo | 6# [A]<sup>8</sup> B-L<sup>8</sup> M<sup>6</su... | 1787 | 1789 | a::91.493.8000:1.13 | Brugge | |
| 6 | c:stcv:12840621 | octavo | 7# [A]<sup>8</sup> B-M<sup>8</sup> N<sup>4</su... | 1787 | 1789 | a::91.493.8000:1.13 | Brugge | |
| 7 | c:stcv:12840621 | octavo | 8# [A]<sup>8</sup> B-2A<sup>8</sup> (2A8 blank) | 1787 | 1789 | a::91.493.8000:1.13 | Brugge | |
| 8 | c:stcv:12840621 | octavo | 9# [A]<sup>8</sup> B-Z<sup>8</sup> 2A<sup>6</sup> | 1787 | 1789 | a::91.493.8000:1.13 | Brugge | |
| 9 | c:stcv:12840621 | octavo | 10# [A]<sup>8</sup> B-Z<sup>8</sup> 2A<sup>4</... | 1787 | 1789 | a::91.493.8000:1.13 | Brugge | |
| 10 | c:stcv:12840621 | octavo | 11# [A]<sup>8</sup> B-Y<sup>8</sup> | 1787 | 1789 | a::91.493.8000:1.13 | Brugge | |

# B. Counting gatherings

The next step is to count the number of sheets quired inside each other thus making up the gatherings, e.g. A-F$^4$ G² contains 6 gatherings of 4 sheets (quire marks A-F) and one of 2 (quire mark G). We also count gatherings with alternating quirings (e.g. 4/2), the multiple usage of quire marks (e.g. A-F$^4$ ²A-F$^4$) and inserted gatherings, registered with π and χ (e.g. A$^8$ (A5 + $^X$A² B-D$^8$). We write this information directly to a second dataframe and then offer some basic information about the dataframe and render the first ten rows. Next, we also save the data to a .csv file ( `STCV_quiring.csv` ) for browsing.

```python
import re
columns_quires = ['1', '2', '3', '4', '5', '6', '7', '8', '9', '10', '11', '12', '13',
'14', '15', '16', '17', '18', '19', '20', '4/2', '4/6', '4/8', '6/8', '8/4', '8/6', 'do
uble', 'triple', 'quadruple', 'quintuple', 'sextuple', 'septuple', 'pi', 'chi']
alternating_quires = ['<sup>4/2</sup> ', '<sup>4/6</sup> ', '<sup>4/8</sup> ', '<sup>6/
8</sup> ', '<sup>8/4</sup> ', '<sup>8/6</sup> ']
inserted_gatherings = ['<sup>π</sup>', '<sup>χ</sup>']
results = []
for i in df1.iloc[:,2]:
    collation = str(i) + ' ' #whitespace added because of the issue described below und
er 'normal quires'
    countofgatherings = []
    ##normal quires
    for i in range(1, 21):
        number = '<sup>' + str(i) + '</sup> '
        #gatherings counted with '</sup> ', i.e. with added whitespace, to avoid counti
ng doubles etc. (e.g. <sup>2</sup>A)
        check = number in str(collation)
        if check == False:
            number_value = 0
        else:
            number_value = 1
        countofgatherings.append(number_value)
    ##alternating quires
    for i in alternating_quires:
        check2 = i in str(collation)
        if check2 == False:
            alternating_value = 0
        else:
            alternating_value = 1
        countofgatherings.append(alternating_value)
    ##multiple quire marks
    for i in range (2, 8):
        multiple = re.compile(r'<sup>' + str(i) + '</sup>\S')
        check3 = multiple.search(str(collation))
        if check3 == None:
            multiple_value = 0
        else:
            multiple_value = 1
        countofgatherings.append(multiple_value)
    ##inserted gatherings
    for i in inserted_gatherings:
        check4 = i in str(collation)
        if check4 == False:
            inserted_value = 0
        else:
            inserted_value = 1
        countofgatherings.append(inserted_value)
    results.append(countofgatherings)
df2 = pd.DataFrame(results, columns=columns_quires)
df2.to_csv('STCV_quiring.csv')
display(df2.info(), df2[0:10])
```

```
<class 'pandas.core.frame.DataFrame'>
RangeIndex: 28292 entries, 0 to 28291
Data columns (total 34 columns):
1            28292 non-null int64
2            28292 non-null int64
3            28292 non-null int64
4            28292 non-null int64
5            28292 non-null int64
6            28292 non-null int64
7            28292 non-null int64
8            28292 non-null int64
9            28292 non-null int64
10           28292 non-null int64
11           28292 non-null int64
12           28292 non-null int64
13           28292 non-null int64
14           28292 non-null int64
15           28292 non-null int64
16           28292 non-null int64
17           28292 non-null int64
18           28292 non-null int64
19           28292 non-null int64
20           28292 non-null int64
4/2          28292 non-null int64
4/6          28292 non-null int64
4/8          28292 non-null int64
6/8          28292 non-null int64
8/4          28292 non-null int64
8/6          28292 non-null int64
double       28292 non-null int64
triple       28292 non-null int64
quadruple    28292 non-null int64
quintuple    28292 non-null int64
sextuple     28292 non-null int64
septuple     28292 non-null int64
pi           28292 non-null int64
chi          28292 non-null int64
dtypes: int64(34)
memory usage: 7.3 MB

None
```

|   | 1 | 2 | 3 | 4 | 5 | 6 | 7 | 8 | 9 | 10 | ... | 8/4 | 8/6 | double | triple | quadruple | quintuple | sextuple |
|---|---|---|---|---|---|---|---|---|---|----|-----|-----|-----|--------|--------|-----------|-----------|----------|
| 0 | 0 | 0 | 0 | 1 | 0 | 1 | 0 | 1 | 0 | 0 | ... | 0 | 0 | 0 | 0 | 0 | 0 | 0 |
| 1 | 0 | 0 | 0 | 1 | 0 | 0 | 0 | 1 | 0 | 0 | ... | 0 | 0 | 0 | 0 | 0 | 0 | 0 |
| 2 | 0 | 0 | 0 | 0 | 0 | 0 | 0 | 1 | 0 | 0 | ... | 0 | 0 | 0 | 0 | 0 | 0 | 0 |
| 3 | 0 | 0 | 0 | 1 | 0 | 0 | 0 | 1 | 0 | 0 | ... | 0 | 0 | 0 | 0 | 0 | 0 | 0 |
| 4 | 0 | 0 | 0 | 0 | 0 | 1 | 0 | 1 | 0 | 0 | ... | 0 | 0 | 0 | 0 | 0 | 0 | 0 |
| 5 | 0 | 0 | 0 | 0 | 0 | 1 | 0 | 1 | 0 | 0 | ... | 0 | 0 | 0 | 0 | 0 | 0 | 0 |
| 6 | 0 | 0 | 0 | 1 | 0 | 0 | 0 | 1 | 0 | 0 | ... | 0 | 0 | 0 | 0 | 0 | 0 | 0 |
| 7 | 0 | 0 | 0 | 0 | 0 | 0 | 0 | 1 | 0 | 0 | ... | 0 | 0 | 0 | 0 | 0 | 0 | 0 |
| 8 | 0 | 0 | 0 | 0 | 0 | 1 | 0 | 1 | 0 | 0 | ... | 0 | 0 | 0 | 0 | 0 | 0 | 0 |
| 9 | 0 | 0 | 0 | 1 | 0 | 0 | 0 | 1 | 0 | 0 | ... | 0 | 0 | 0 | 0 | 0 | 0 | 0 |

10 rows × 34 columns

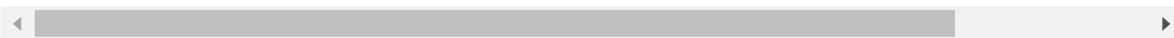

We can then combine both dataframes in a third dataframe. We offer some basic information about it and render the first ten rows. We then also save the data to a .csv file ( STCV_all_quiring.csv ) for browsing.

```python
df3 = pd.concat([df1, df2], axis=1, sort=False)
display(df3.info(), df3[0:11])
df3.to_csv('STCV_all_quiring.csv') #check nog eens op deze csv paar random records of a
lle data echt klopt!!
```

```
<class 'pandas.core.frame.DataFrame'>
RangeIndex: 28292 entries, 0 to 28291
Data columns (total 43 columns):
identifier       28292 non-null object
format           28292 non-null object
collation        28292 non-null object
year1            28292 non-null object
year2            28292 non-null object
place_code       28292 non-null object
place_name       28292 non-null object
publisher_code   28292 non-null object
publisher_name   28292 non-null object
1                28292 non-null int64
2                28292 non-null int64
3                28292 non-null int64
4                28292 non-null int64
5                28292 non-null int64
6                28292 non-null int64
7                28292 non-null int64
8                28292 non-null int64
9                28292 non-null int64
10               28292 non-null int64
11               28292 non-null int64
12               28292 non-null int64
13               28292 non-null int64
14               28292 non-null int64
15               28292 non-null int64
16               28292 non-null int64
17               28292 non-null int64
18               28292 non-null int64
19               28292 non-null int64
20               28292 non-null int64
4/2              28292 non-null int64
4/6              28292 non-null int64
4/8              28292 non-null int64
6/8              28292 non-null int64
8/4              28292 non-null int64
8/6              28292 non-null int64
double           28292 non-null int64
triple           28292 non-null int64
quadruple        28292 non-null int64
quintuple        28292 non-null int64
sextuple         28292 non-null int64
septuple         28292 non-null int64
pi               28292 non-null int64
chi              28292 non-null int64
dtypes: int64(34), object(9)
memory usage: 8.3+ MB

None
```

|  | identifier | format | collation | year1 | year2 | place_code | place_name | p |
|---|---|---|---|---|---|---|---|---|
| 0 | c:stcv:12840621 | octavo | 1# π<sup>4</sup> A-Z<sup>8</sup> 2A<sup>6</sup> | 1787 | 1789 | a::91.493.8000:1.13 | Brugge |  |
| 1 | c:stcv:12840621 | octavo | 2# [A]<sup>8</sup> B-2A<sup>8</sup> 2B<sup>4</... | 1787 | 1789 | a::91.493.8000:1.13 | Brugge |  |
| 2 | c:stcv:12840621 | octavo | 3# [A]<sup>8</sup> B-Z<sup>8</sup> (Z8 blank) | 1787 | 1789 | a::91.493.8000:1.13 | Brugge |  |
| 3 | c:stcv:12840621 | octavo | 4# [A]<sup>8</sup> B-L<sup>8</sup> M<sup>4</su... | 1787 | 1789 | a::91.493.8000:1.13 | Brugge |  |
| 4 | c:stcv:12840621 | octavo | 5# [A]<sup>8</sup> B-L<sup>8</sup> M<sup>6</su... | 1787 | 1789 | a::91.493.8000:1.13 | Brugge |  |
| 5 | c:stcv:12840621 | octavo | 6# [A]<sup>8</sup> B-L<sup>8</sup> M<sup>6</su... | 1787 | 1789 | a::91.493.8000:1.13 | Brugge |  |
| 6 | c:stcv:12840621 | octavo | 7# [A]<sup>8</sup> B-M<sup>8</sup> N<sup>4</su... | 1787 | 1789 | a::91.493.8000:1.13 | Brugge |  |
| 7 | c:stcv:12840621 | octavo | 8# [A]<sup>8</sup> B-2A<sup>8</sup> (2A8 blank) | 1787 | 1789 | a::91.493.8000:1.13 | Brugge |  |
| 8 | c:stcv:12840621 | octavo | 9# [A]<sup>8</sup> B-Z<sup>8</sup> 2A<sup>6</sup> | 1787 | 1789 | a::91.493.8000:1.13 | Brugge |  |
| 9 | c:stcv:12840621 | octavo | 10# [A]<sup>8</sup> B-Z<sup>8</sup> 2A<sup>4</... | 1787 | 1789 | a::91.493.8000:1.13 | Brugge |  |
| 10 | c:stcv:12840621 | octavo | 11# [A]<sup>8</sup> B-Y<sup>8</sup> | 1787 | 1789 | a::91.493.8000:1.13 | Brugge |  |

11 rows × 43 columns

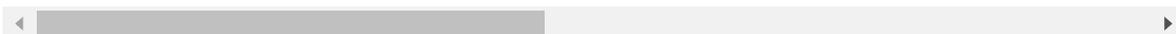

# C. Analysing and visualizing gatherings

After collecting the data and counting the gatherings, we can now analyse and visualize the results of this operation. For instance, we can make a sum of all quiring practices (quires of ones, twos, threes, ...) and plot the result.

In [31]:

```
#We keep the code for this example and the next explicit and verbose to clearly show th
e method.
#Therefore the same code snippets will be repeated several times instead of defining cu
stom functions.
df4 = pd.DataFrame(columns=columns_quires)
sum_data = []
for i in columns_quires:
        sum_data.append(df3[i].sum())
df4.loc[0] = sum_data
display(df4)

import matplotlib.pyplot as plt
%matplotlib inline
plt.figure()
df4.iloc[0].plot(kind='bar', title='General')
```

|   | 1 | 2 | 3 | 4 | 5 | 6 | 7 | 8 | 9 | 10 | ... | 8/4 | 8/6 | double | triple | quadruple |
|---|---|---|---|---|---|---|---|---|---|----|-----|-----|-----|--------|--------|-----------|
| 0 | 27 | 8004 | 1 | 12920 | 2 | 4930 | 2 | 9808 | 2 | 468 | ... | 199 | 1 | 1417 | 235 | 322 |

1 rows × 34 columns

Out[31]:

\<matplotlib.axes._subplots.AxesSubplot at 0x113acfb0\>

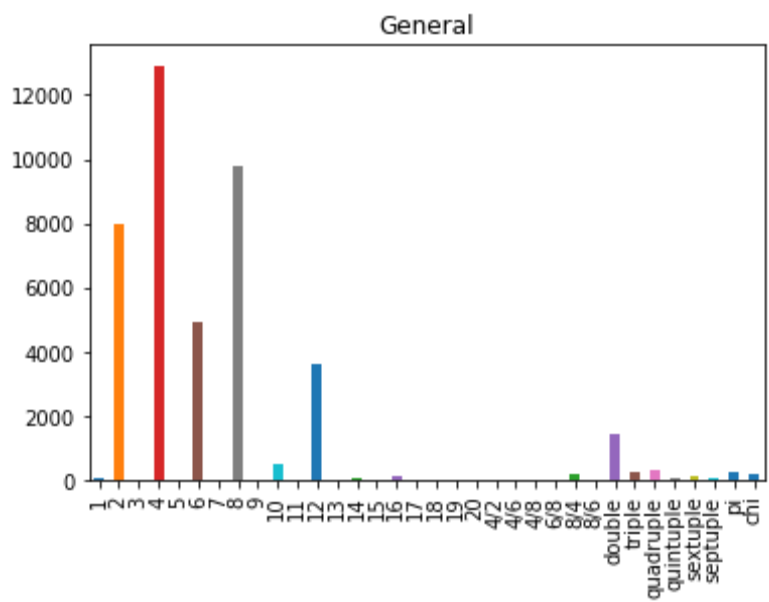

Or, we can also select a certain bibliographic format, e.g. `quarto`, to see how this compares to quiring practices in general.

```python
df_quarto = df3.loc[df3['format'] == 'quarto']
df4 = pd.DataFrame(columns=columns_quires)
sum_data = []
for i in columns_quires:
        sum_data.append(df_quarto[i].sum())
df4.loc[0] = sum_data
display(df4)

import matplotlib.pyplot as plt
%matplotlib inline
plt.figure()
df4.iloc[0].plot(kind='bar', title='Quarto')
```

| | 1 | 2 | 3 | 4 | 5 | 6 | 7 | 8 | 9 | 10 | ... | 8/4 | 8/6 | double | triple | quadruple | quintupl |
|---|---|---|---|---|---|---|---|---|---|---|---|---|---|---|---|---|---|
| 0 | 2 | 2834 | 0 | 5230 | 0 | 654 | 0 | 470 | 0 | 52 | ... | 2 | 1 | 272 | 52 | 111 | 1 |

1 rows × 34 columns

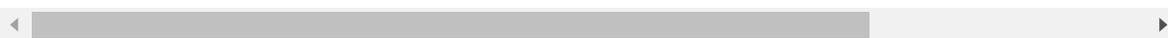

```
<matplotlib.axes._subplots.AxesSubplot at 0x10b9fe50>
```

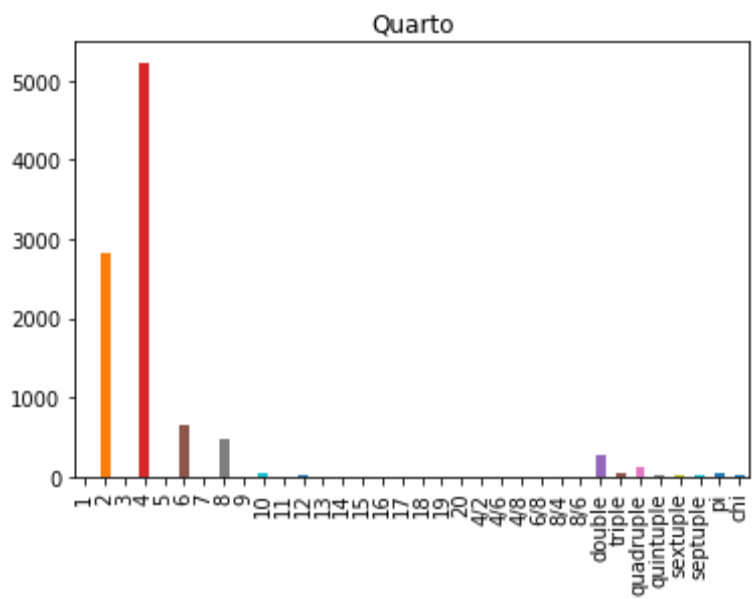

Or, we could have a look at sixteenth-century quiring practices.

```python
df_15 = df3.loc[pd.to_numeric(df3['year1']).isin(range(1500,1600))]
df4 = pd.DataFrame(columns=columns_quires)
sum_data = []
for i in columns_quires:
        sum_data.append(df_15[i].sum())
df4.loc[0] = sum_data
display(df4)

import matplotlib.pyplot as plt
%matplotlib inline
plt.figure()
df4.iloc[0].plot(kind='bar', title='Sixteenth')
```

|   | 1 | 2 | 3 | 4 | 5 | 6 | 7 | 8 | 9 | 10 | ... | 8/4 | 8/6 | double | triple | quadruple | quintupl |
|---|---|---|---|---|---|---|---|---|---|---|---|---|---|---|---|---|---|
| **0** | 2 | 162 | 0 | 1017 | 0 | 362 | 0 | 1188 | 0 | 74 | ... | 18 | 1 | 158 | 11 | 20 | |

1 rows × 34 columns

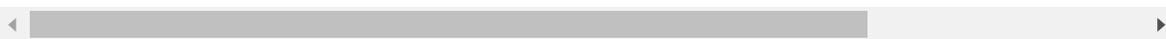

```
<matplotlib.axes._subplots.AxesSubplot at 0x10620990>
```

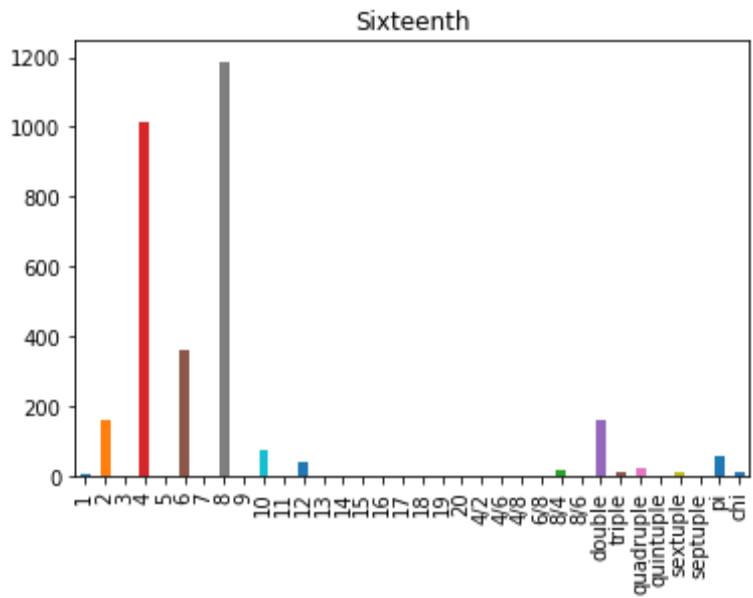

Or, at sixteenth-century quartos in particular.

```python
df_15 = df3.loc[pd.to_numeric(df3['year1']).isin(range(1500,1600))]
df_15_quarto = df_15.loc[df_15['format'] == 'quarto']
df4 = pd.DataFrame(columns=columns_quires)
sum_data = []
for i in columns_quires:
        sum_data.append(df_15_quarto[i].sum())
df4.loc[0] = sum_data
display(df4)

import matplotlib.pyplot as plt
%matplotlib inline
plt.figure()
df4.iloc[0].plot(kind='bar', title='16-c. quarto')
```

| | 1 | 2 | 3 | 4 | 5 | 6 | 7 | 8 | 9 | 10 | ... | 8/4 | 8/6 | double | triple | quadruple | quintuple | se |
|---|---|---|---|---|---|---|---|---|---|---|---|---|---|---|---|---|---|---|
| 0 | 1 | 82 | 0 | 430 | 0 | 97 | 0 | 73 | 0 | 10 | ... | 1 | 1 | 29 | 3 | 8 | 0 | |

1 rows × 34 columns

```
<matplotlib.axes._subplots.AxesSubplot at 0x11939d90>
```

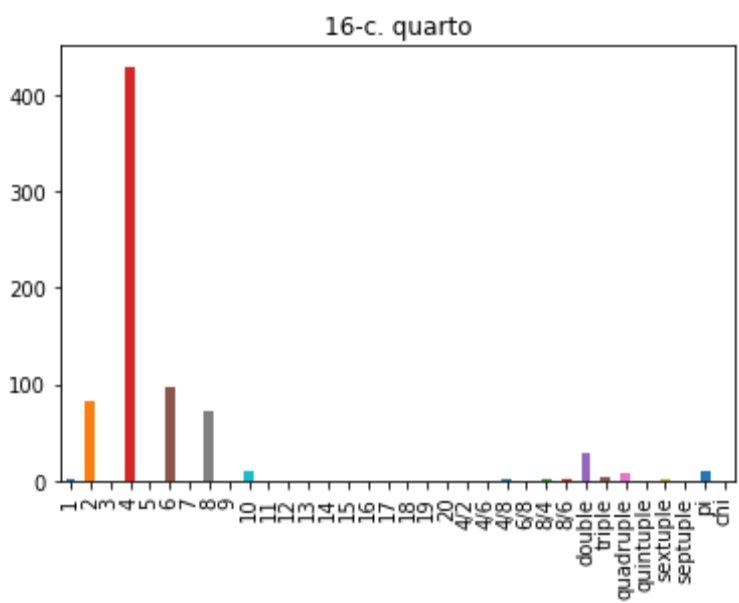

Or, all sixteenth-century quartos printed in Antwerp.

```python
Antwerp = ['a::91.493.2000:1.43', 'a::91.493.2000:1.42' 'a::91.493.2000:1.28' 'a::91.49
3.2000:1.15' 'a::91.493.2000:1 ','a::91.493.2000:1.11 ','a::91.493.2000:1.44 ','a::91.4
93.2000:1.68 ','a::91.493.2000:1.119 ','a::91.493.2000:1.45 ','a::91.493.2000:1.71 ',
'a::91.493.2000:1.35 ','a::91.493.2000:1.46 ','a::91.493.2000:1.118 ','a::91.493.2000:
1.171 ','a::91.493.2000:1.37 ','a::91.493.2000:1.1 ','a::91.493.2000:1.19 ','a::91.493.
2000:1.54 ','a::91.493.2000:1.63 ','a::91.493.2000:1.135 ','a::91.493.2000:1.49 ','a::9
1.493.2000:1.111 ','a::91.493.2000:1.50 ','a::91.493.2000:1.41 ','a::91.493.2000:1.51 '
,'a::91.493.2000:1.52 ','a::91.493.2000:1.53 ','a::91.493.2000:1.101 ','a::91.493.2000:
1.58 ','a::91.493.2000:1.16 ','a::91.493.2000:1.26 ','a::91.493.2000:1.59 ','a::91.493.
2000:1:N ','a::91.493.2000:1.22 ','a::91.493.2000:1.61 ','a::91.493.2000:1.121 ','a::9
1.493.2000:1.122 ','a::91.493.2000:1.137 ','a::91.493.2000:1.48 ','a::91.493.2000:1.123
 ','a::91.493.2000:1.47 ','a::91.493.2000:1.62 ','a::91.493.2000:1.127 ','a::91.493.200
0:1.128 ','a::91.493.2000:1.69 ','a::91.493.2000:1.90 ','a::91.493.2000:1.130 ','a::91.
493.2000:1.70 ','a::91.493.2000:1.134 ','a::91.493.2000:1.138 ','a::91.493.2000:1.139 '
,'a::91.493.2000:1.140 ','a::91.493.2000:1.141 ','a::91.493.2000:1.142 ','a::91.493.200
0:1.143 ','a::91.493.2000:1.76 ','a::91.493.2000:1.78 ','a::91.493.2000:1.108 ','a::91.
493.2000:1.146 ','a::91.493.2000:1.147 ','a::91.493.2000:1.40 ','a::91.493.2000:1.151 '
,'a::91.493.2000:1.152 ','a::91.493.2000:1.104 ','a::91.493.2000:1.153 ','a::91.493.200
0:1.155 ','a::91.493.2000:1.154 ','a::91.493.2000:1.20 ','a::91.493.2000:1.25 ','a::91.
493.2000:1.100 ','a::91.493.2000:1.156 ','a::91.493.2000:1.9 ','a::91.493.2000:1.103 ',
'a::91.493.2000:1.158 ','a::91.493.2000:1.6 ','a::91.493.2000:1.83 ','a::91.493.2000:1.
160 ','a::91.493.2000:1.24 ','a::91.493.2000:1.162 ','a::91.493.2000:1.77 ','a::91.493.
2000:1.163 ','a::91.493.2000:1.79 ','a::91.493.2000:1.164 ','a::91.493.2000:1.99 ','a::
91.493.2000:1.165 ','a::91.493.2000:1.3 ','a::91.493.2000:1.67 ','a::91.493.2000:1.166
 ','a::91.493.2000:1.167 ','a::91.493.2000:1.75 ','a::91.493.2000:1.74 ','a::91.493.200
0:1.91 ','a::91.493.2000:1.172 ','a::91.493.2000:1.82 ','a::91.493.2000:1.84 ','a::91.4
93.2000:1.2 ','a::91.493.2000:1.175 ','a::91.493.2000:1.176 ','a::91.493.2000:1.12 ',
'a::91.493.2000:1.113 ','a::91.493.2000:1.179 ','a::91.493.2000:1.5 ','a::91.493.2000:
1.65 ','a::91.493.2000:1.55 ','a::91.493.2000:1.86 ','a::91.493.2000:1.181 ','a::91.49
3.2000:1:F ','a::91.493.2000:1.93 ','a::91.493.2000:1.94 ','a::91.493.2000:1.96 ','a::9
1.493.2000:1.183 ','a::91.493.2000:1.184 ','a::91.493.2000:1.14 ','a::91.493.2000:1.191
 ','a::91.493.2000:1.186 ','a::91.493.2000:1.98 ','a::91.493.2000:1.192 ','a::91.493.20
00:1.148 ','a::91.493.2000:1.193 ','a::91.493.2000:1.149 ','a::91.493.2000:1.194 ','a::
91.493.2000:1.195 ','a::91.493.2000:1.196 ','a::91.493.2000:1.197 ','a::91.493.2000:1.1
98 ','a::91.493.2000:1.199 ','a::91.493.2000:1.200 ','a::91.493.2000:1.102 ','a::91.49
3.2000:1.202 ','a::91.493.2000:1.203 ','a::91.493.2000:1.204 ','a::91.493.2000:1.105 ',
'a::91.493.2000:1.205 ','a::91.493.2000:1.206 ','a::91.493.2000:1.107 ','a::91.493.200
0:1.109 ','a::91.493.2000:1.112 ','a::91.493.2000:1.10 ','a::91.493.2000:1.115 ','a::9
1.493.2000:1.116 ','a::91.493.2000:1.85']
df_15 = df3.loc[pd.to_numeric(df3['year1']).isin(range(1500,1600))]
df_15_quarto = df_15.loc[df_15['format'] == 'quarto']
df_15_quarto_Antwerp = df_15_quarto.loc[df_15['place_code'].isin(Antwerp)]
df4 = pd.DataFrame(columns=columns_quires)
sum_data = []
for i in columns_quires:
        sum_data.append(df_15_quarto_Antwerp[i].sum())
df4.loc[0] = sum_data
display(df4)

import matplotlib.pyplot as plt
%matplotlib inline
plt.figure()
df4.iloc[0].plot(kind='bar', title='16-c. quarto Antwerp')
```

|  | 1 | 2 | 3 | 4 | 5 | 6 | 7 | 8 | 9 | 10 | ... | 8/4 | 8/6 | double | triple | quadruple | quintuple | sextupl |
|---|---|---|---|---|---|---|---|---|---|---|---|---|---|---|---|---|---|---|
| **0** | 0 | 3 | 0 | 19 | 0 | 5 | 0 | 6 | 0 | 2 | ... | 0 | 0 | 1 | 0 | 0 | 0 |  |

1 rows × 34 columns

<matplotlib.axes._subplots.AxesSubplot at 0x4305070>

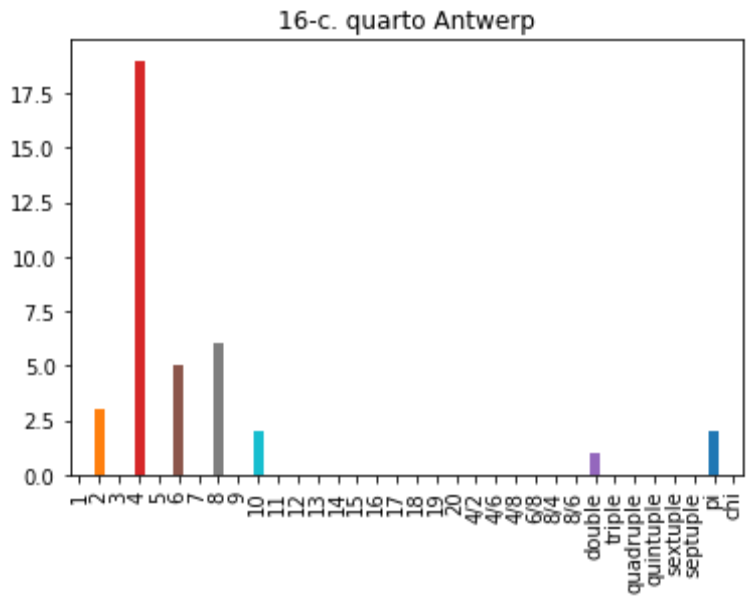

Finally, we can use stacked bars to plot some graphs that allow to compare quiring practices over different formats and different centuries. We also save these dataframes to .csv files for browsing ( `STCV_formats_vs_quiring.csv` and `STCV_centuries_vs_quiring.csv` )

```python
##formats
sum_all = []
formats = ['plano', 'folio', 'quarto', 'octavo', 'duodecimo', 'sextodecimo', 'octodecim
o', 'vicesimoquarto', 'tricesimosecundo']
for i in formats:
    df_new = df3.loc[df3['format'] == i]
    sum_data = []
    for k in columns_quires:
        sum_data.append(df_new[k].sum())
    sum_all.append(sum_data)
df4 = pd.DataFrame(sum_all, columns=columns_quires, index=formats)
display(df4)
df4.to_csv('STCV_formats_vs_quiring.csv')

import matplotlib.pyplot as plt
%matplotlib inline
plt.figure()
df4.plot(kind='bar', stacked='True', title='Format vs quiring practices', legend=False)
df4 = pd.DataFrame(columns=columns_quires, index=formats)
df4.plot(title='Legend')
```

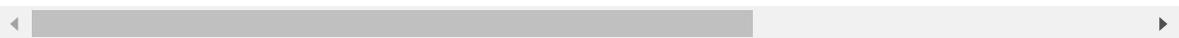

| | 1 | 2 | 3 | 4 | 5 | 6 | 7 | 8 | 9 | 10 | ... | 8/4 | 8/6 | double | triple |
|---|---|---|---|---|---|---|---|---|---|---|---|---|---|---|---|
| **plano** | 2 | 2 | 0 | 5 | 0 | 2 | 0 | 0 | 0 | 0 | ... | 0 | 0 | 0 | 0 |
| **folio** | 6 | 2130 | 0 | 1916 | 1 | 1634 | 2 | 741 | 2 | 119 | ... | 0 | 0 | 543 | 117 |
| **quarto** | 2 | 2834 | 0 | 5230 | 0 | 654 | 0 | 470 | 0 | 52 | ... | 2 | 1 | 272 | 52 |
| **octavo** | 14 | 2300 | 0 | 4705 | 1 | 796 | 0 | 7284 | 0 | 129 | ... | 1 | 0 | 399 | 42 |
| **duodecimo** | 3 | 671 | 1 | 908 | 0 | 1716 | 0 | 628 | 0 | 165 | ... | 185 | 0 | 166 | 24 |
| **sextodecimo** | 0 | 16 | 0 | 50 | 0 | 16 | 0 | 380 | 0 | 1 | ... | 2 | 0 | 13 | 0 |
| **octodecimo** | 0 | 30 | 0 | 33 | 0 | 76 | 0 | 21 | 0 | 1 | ... | 0 | 0 | 8 | 0 |
| **vicesimoquarto** | 0 | 19 | 0 | 70 | 0 | 34 | 0 | 266 | 0 | 0 | ... | 9 | 0 | 15 | 0 |
| **tricesimosecundo** | 0 | 2 | 0 | 3 | 0 | 2 | 0 | 18 | 0 | 1 | ... | 0 | 0 | 1 | 0 |

9 rows × 34 columns

Out[36]:

<matplotlib.axes._subplots.AxesSubplot at 0xefb6c30>

<Figure size 432x288 with 0 Axes>

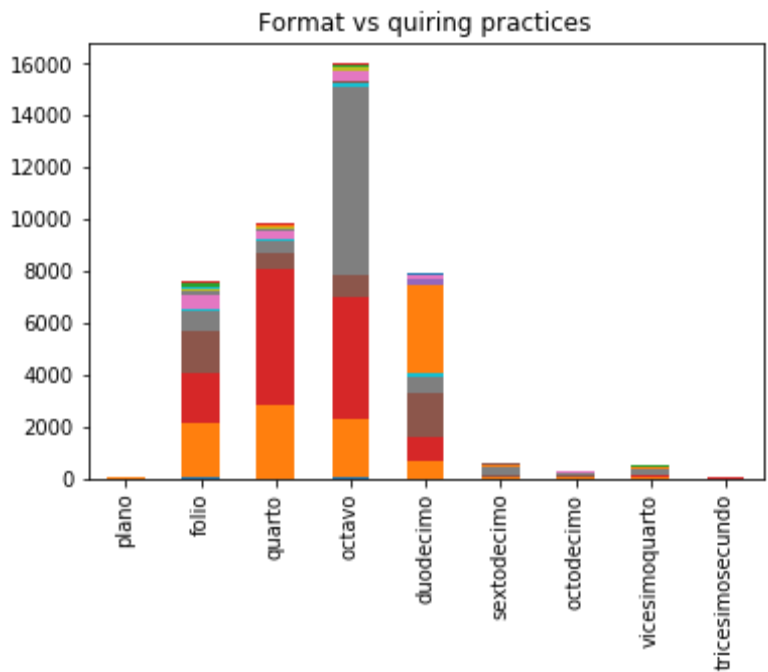

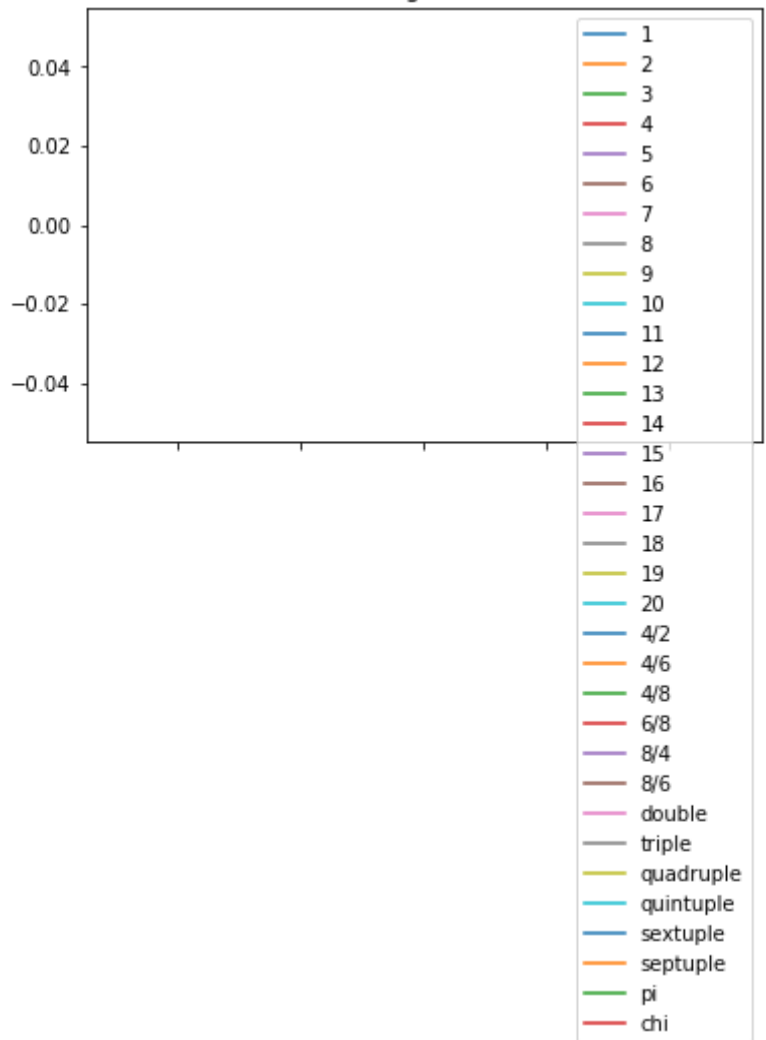

```python
##centuries
sum_all = []
centuries = [range(1400,1499), range(1500,1599), range(1600,1699), range(1700,1799)]
for i in centuries:
    df_newer = df3.loc[pd.to_numeric(df3['year1']).isin(i)]
    sum_data = []
    for k in columns_quires:
        sum_data.append(df_newer[k].sum())
    sum_all.append(sum_data)
df4 = pd.DataFrame(sum_all, columns=columns_quires)
display(df4)
df4.to_csv('STCV_centuries_vs_quiring.csv')

import matplotlib.pyplot as plt
%matplotlib inline
plt.figure()
df4.plot(kind='bar', stacked='True', title='15th, 16th, 17th, 18th Century vs quiring p
ractices', legend=False)
df4 = pd.DataFrame(columns=columns_quires, index=formats)
df4.plot(title='Legend')
```

| | 1 | 2 | 3 | 4 | 5 | 6 | 7 | 8 | 9 | 10 | ... | 8/4 | 8/6 | double | triple | quadruple | qui |
|---|---|---|---|---|---|---|---|---|---|---|---|---|---|---|---|---|---|
| **0** | 0 | 1 | 0 | 22 | 0 | 54 | 2 | 78 | 2 | 28 | ... | 1 | 0 | 1 | 0 | 0 | |
| **1** | 1 | 160 | 0 | 997 | 0 | 350 | 0 | 1178 | 0 | 71 | ... | 18 | 1 | 152 | 9 | 19 | |
| **2** | 3 | 3267 | 0 | 7015 | 1 | 2714 | 0 | 4729 | 0 | 184 | ... | 78 | 0 | 760 | 117 | 154 | |
| **3** | 22 | 4330 | 1 | 4627 | 0 | 1680 | 0 | 3595 | 0 | 177 | ... | 98 | 0 | 476 | 100 | 144 | |

4 rows × 34 columns

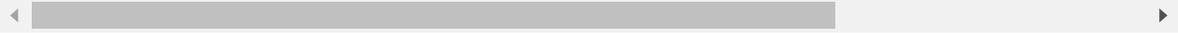

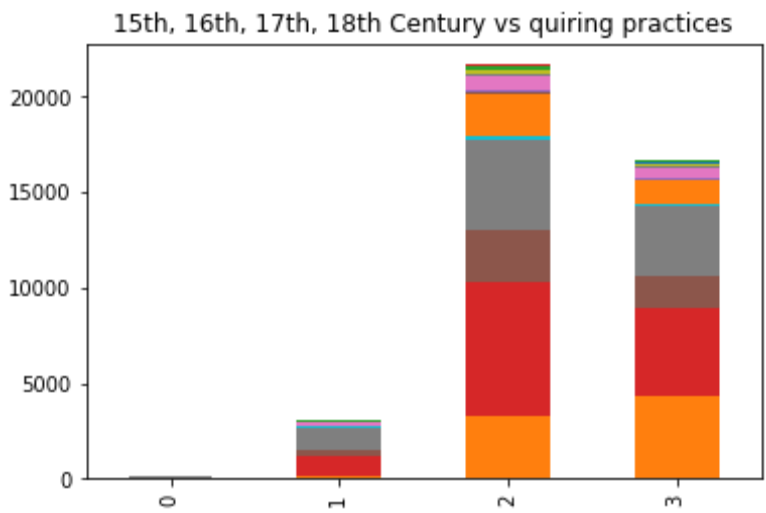

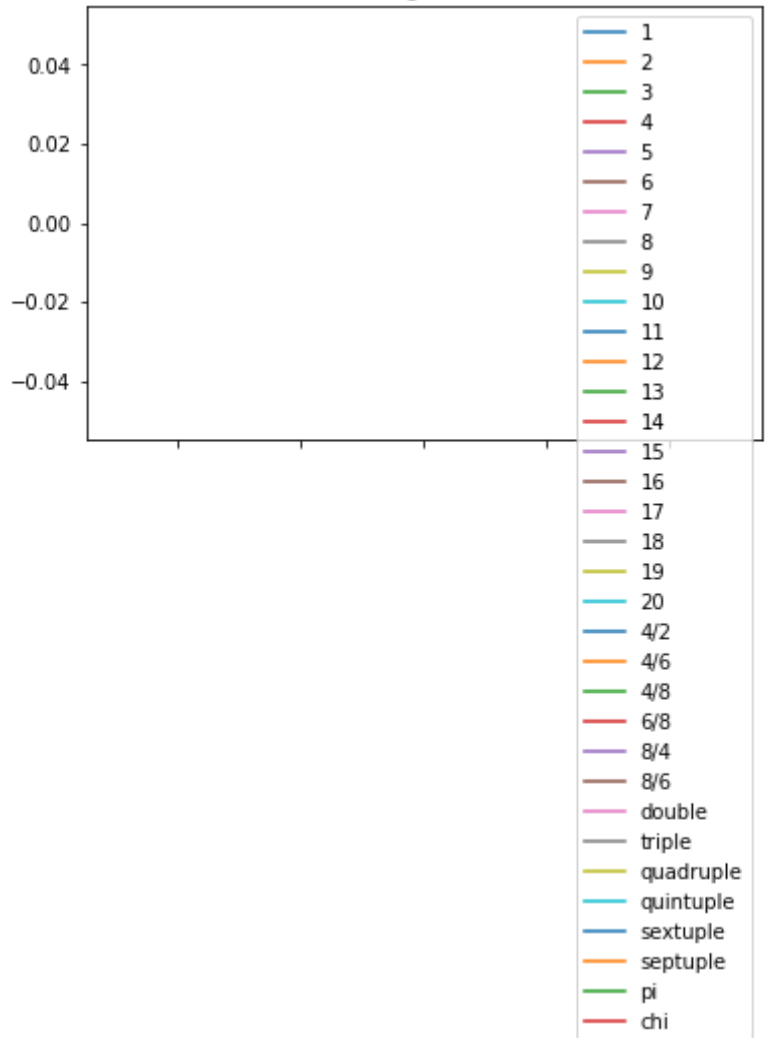

```python
##evolution
sum_all = []
formats_standard = ['folio', 'quarto', 'octavo', 'duodecimo']
centuries = [range(1400,1499), range(1500,1599), range(1600,1699), range(1700,1799)]
for i in formats_standard:
    df_newer = df3.loc[df3['format'] == i]
    for j in centuries:
        df_newest = df_newer.loc[pd.to_numeric(df_newer['year1']).isin(j)]
        sum_data = []
        for k in columns_quires:
            sum_data.append(df_newest[k].sum())
        sum_all.append(sum_data)

indices = ['15th-c folio', '16th-c folio', '17th-c folio', '18th-c folio', '15th-c quar
to', '16th-c quarto', '17th-c quarto', '18th-c quarto', '15th-c octavo', '16th-c octav
o', '17th-c octavo', '18th-c octavo', '15th-c duodecimo', '16th-c duodecimo', '17th-c d
uodecimo', '18th-c duodecimo']
df4 = pd.DataFrame(sum_all, index=indices)
display(df4)
df4.to_csv('STCV_centuries_vs_quiring.csv')

import matplotlib.pyplot as plt
%matplotlib inline
plt.figure()
df4.plot(kind='bar', stacked='True', title='15th, 16th, 17th, 18th folios, quartos, oct
avos and duodecimos', legend=False)
df4 = pd.DataFrame(columns=columns_quires, index=formats)
df4.plot(title='Legend')
```

|  | 0 | 1 | 2 | 3 | 4 | 5 | 6 | 7 | 8 | 9 | ... | 24 | 25 | 26 | 27 | 28 | 29 | 30 |
|---|---|---|---|---|---|---|---|---|---|---|---|---|---|---|---|---|---|---|
| **15th-c folio** | 0 | 1 | 0 | 7 | 0 | 22 | 2 | 41 | 2 | 23 | ... | 0 | 0 | 1 | 0 | 0 | 0 | 0 |
| **16th-c folio** | 0 | 56 | 0 | 177 | 0 | 211 | 0 | 119 | 0 | 34 | ... | 0 | 0 | 74 | 5 | 4 | 0 | 7 |
| **17th-c folio** | 2 | 785 | 0 | 1152 | 1 | 1183 | 0 | 502 | 0 | 54 | ... | 0 | 0 | 323 | 65 | 46 | 8 | 28 |
| **18th-c folio** | 3 | 1257 | 0 | 540 | 0 | 195 | 0 | 69 | 0 | 4 | ... | 0 | 0 | 130 | 40 | 38 | 25 | 20 |
| **15th-c quarto** | 0 | 0 | 0 | 11 | 0 | 28 | 0 | 27 | 0 | 3 | ... | 1 | 0 | 0 | 0 | 0 | 0 | 0 |
| **16th-c quarto** | 1 | 81 | 0 | 420 | 0 | 97 | 0 | 73 | 0 | 10 | ... | 1 | 1 | 28 | 3 | 7 | 0 | 1 |
| **17th-c quarto** | 0 | 1619 | 0 | 3620 | 0 | 328 | 0 | 208 | 0 | 24 | ... | 0 | 0 | 175 | 32 | 56 | 2 | 10 |
| **18th-c quarto** | 1 | 1053 | 0 | 1090 | 0 | 184 | 0 | 151 | 0 | 15 | ... | 0 | 0 | 67 | 16 | 46 | 10 | 11 |
| **15th-c octavo** | 0 | 0 | 0 | 4 | 0 | 4 | 0 | 8 | 0 | 2 | ... | 0 | 0 | 0 | 0 | 0 | 0 | 0 |
| **16th-c octavo** | 0 | 21 | 0 | 377 | 0 | 24 | 0 | 841 | 0 | 27 | ... | 0 | 0 | 43 | 1 | 8 | 0 | 0 |
| **17th-c octavo** | 1 | 548 | 0 | 1665 | 0 | 192 | 0 | 3206 | 0 | 29 | ... | 0 | 0 | 162 | 13 | 39 | 1 | 4 |
| **18th-c octavo** | 13 | 1620 | 0 | 2551 | 0 | 520 | 0 | 3047 | 0 | 71 | ... | 1 | 0 | 189 | 27 | 52 | 7 | 14 |
| **15th-c duodecimo** | 0 | 0 | 0 | 0 | 0 | 0 | 0 | 0 | 0 | 0 | ... | 0 | 0 | 0 | 0 | 0 | 0 | 0 |
| **16th-c duodecimo** | 0 | 0 | 0 | 2 | 0 | 15 | 0 | 8 | 0 | 0 | ... | 17 | 0 | 1 | 0 | 0 | 0 | 0 |
| **17th-c duodecimo** | 0 | 280 | 0 | 470 | 0 | 935 | 0 | 346 | 0 | 75 | ... | 67 | 0 | 82 | 7 | 10 | 2 | 11 |
| **18th-c duodecimo** | 3 | 370 | 1 | 417 | 0 | 738 | 0 | 259 | 0 | 86 | ... | 97 | 0 | 81 | 17 | 8 | 3 | 5 |

16 rows × 34 columns

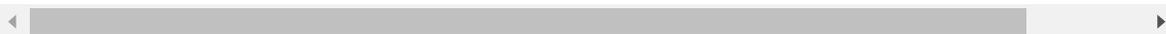

Out[38]:

<matplotlib.axes._subplots.AxesSubplot at 0xd2fac90>

<Figure size 432x288 with 0 Axes>

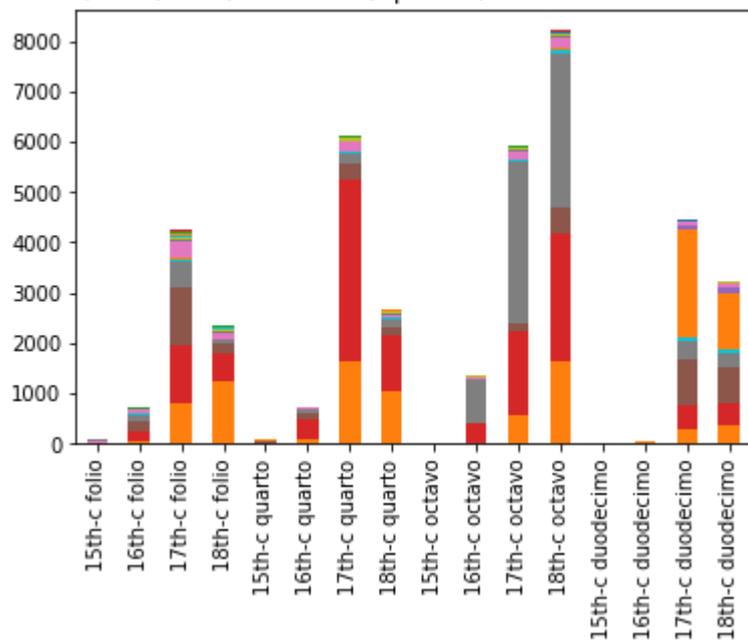

15th, 16th, 17th, 18th folios, quartos, octavos and duodecimos

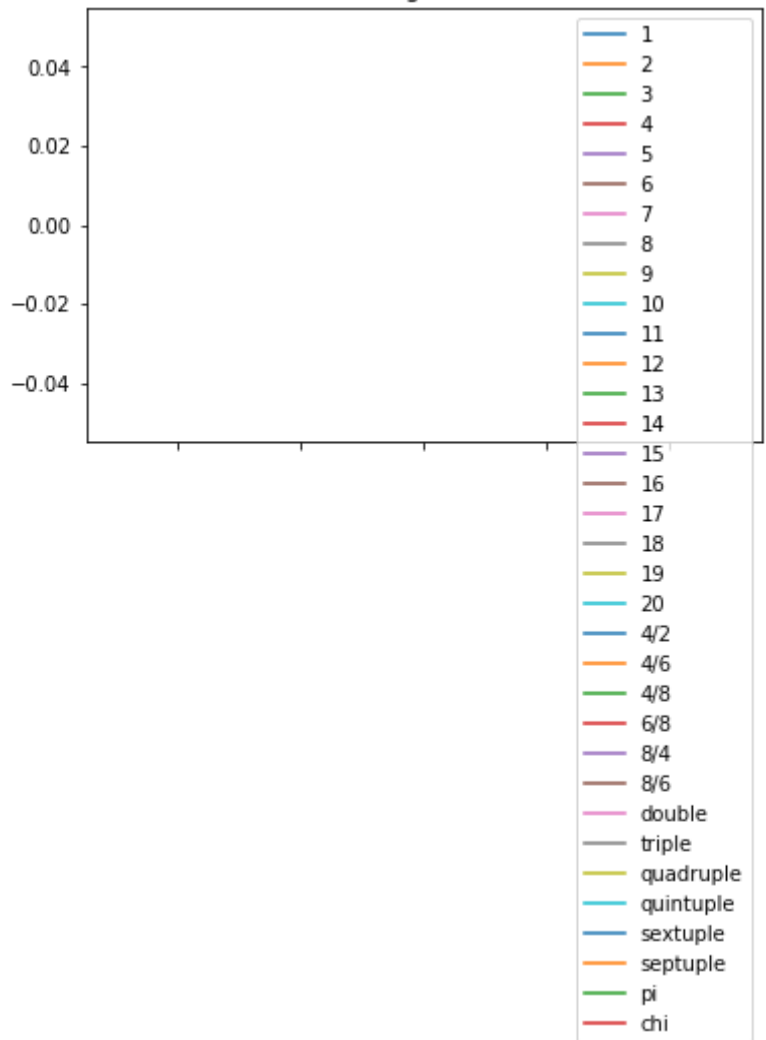

## D. Interpretation

The detailed interpretation of these results is beyond the limited scope of this paper. However, it is clear that the dataset can provide a good start to interpret Early Modern quiring. For instance, Gaskell states the following on the subject:

*In the fifteenth century folio gatherings consisted of up to five sheets; quarto (and occasionally octavo) gatherings might consist of two sheets. Patterns of quiring changed in later periods. Folios were generally gathered in 6s during the sixteenth and seventeenth centuries, but most eighteenth-century folios were gathered by single sheets (i.e. in 2s) despite the extra sewing this entailed. Quarto in 8s remained common in English printing until the seventeenth-century, and was continued in Bible printing until 1800. Octavo gatherings were rarely quired after the fifteenth century.* (GASKELL 2012: 82-83)

From our research, however, we see that Gaskell's assertion on sixteenth- and seventeenth-century folios being generally gathered in 6s does not really hold true (4s are equally popular and 2s and 8s are not that far behind), at least for this corpus of editions printed in Flanders.

```python
df_1516 = df3.loc[pd.to_numeric(df3['year1']).isin(range(1500,1699))]
df_1516_folio = df_1516.loc[df_1516['format'] == 'folio']
df4 = pd.DataFrame(columns=columns_quires)
sum_data = []
for i in columns_quires:
        sum_data.append(df_1516_folio[i].sum())
df4.loc[0] = sum_data
display(df4)

import matplotlib.pyplot as plt
%matplotlib inline
plt.figure()
df4.iloc[0].plot(kind='bar', title='16-c and 17-c. folios')
```

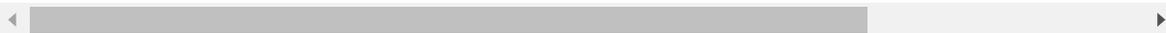

|   | 1 | 2 | 3 | 4 | 5 | 6 | 7 | 8 | 9 | 10 | ... | 8/4 | 8/6 | double | triple | quadruple | quintupl |
|---|---|---|---|---|---|---|---|---|---|----|-----|-----|-----|--------|--------|-----------|----------|
| 0 | 3 | 842 | 0 | 1339 | 1 | 1404 | 0 | 625 | 0 | 91 | ... | 0 | 0 | 402 | 72 | 50 | |

1 rows × 34 columns

```
<matplotlib.axes._subplots.AxesSubplot at 0xeef9f50>
```

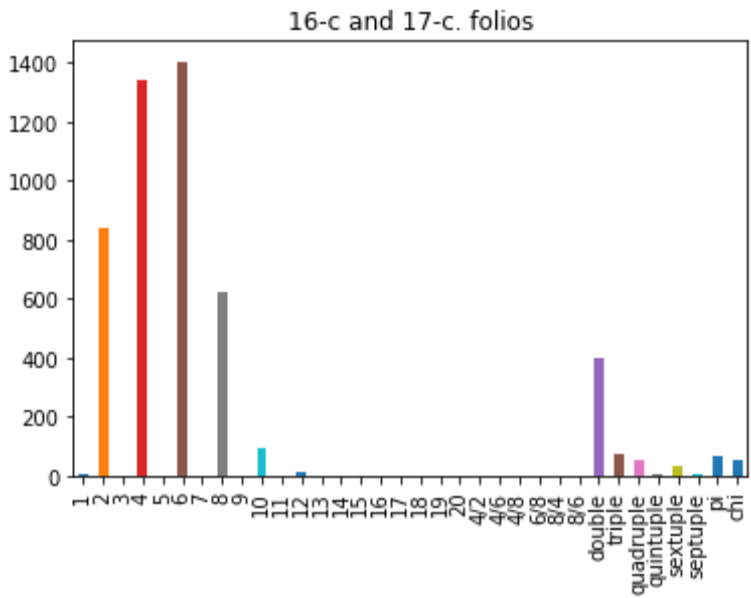

Obviously the kind of datamining as presented in this notebook could serve well to further our understanding of quiring practices, which in the past could only rely on smallish data samples (if any).

Furthermore, in this preliminary research we have only looked at the relation between the collation field, bibliographic format, date of publication and place, yet the other metadata available for the corpus, such as printer, language, subject matter, etcetera, offer a plethora of additional possibilities for interpreting quiring patterns. Moreover, it is important to note that our algorithm only records whether an edition contains a certain quiring or not, whereas it does not count the amount of gatherings that use a certain quiring.

Finally, it is important to note that we can also use this code for other datasets, such as the full Anet dataset of handpress books, which contains more than 17,000 editions of handpress books, since its datastructure is the same as the STCV dataset, which allows for the same SQLite and Pandas queries to be used.